\documentclass[aps,prb,groupedaddress]{revtex4}
\usepackage{graphicx}
\usepackage{amsmath}

\begin{document}

\title{Improved ``Position Squared'' Readout of a Mechanical Resonator in an Optical Cavity Using Degenerate Optical Modes}

\author{J. C. Sankey$^1$, A. M. Jayich$^2$, B. M. Zwickl$^2$, C. Yang$^2$, J. G. E. Harris$^{*,1,2}$}

\affiliation{%
  $^1$Department of Applied Physics, Yale University\\
  $^2$Department of Physics, Yale University\\
  New Haven, CT 06520, USA\\
}

\begin{abstract}
Optomechanical devices in which a flexible SiN membrane is placed inside an optical cavity allow for very high finesse and mechanical quality factor in a single device. They also provide fundamentally new functionality: the cavity detuning can be a quadratic function of membrane position. This enables a measurement of ``position squared'' ($x^2$) and in principle a QND phonon number readout of the membrane. However, the readout achieved using a single transverse cavity mode is not sensitive enough to observe quantum jumps between phonon Fock states. 

Here we demonstrate an $x^2$-sensitivity that is orders of magnitude stronger using two transverse cavity modes that are nearly degenerate. We derive a first-order perturbation theory to describe the interactions between nearly-degenerate cavity modes and achieve good agreement with our measurements using realistic parameters. We also demonstrate theoretically that the $x^2$-coupling should be easily tunable over a wide range.

\end{abstract}


\maketitle

\section{Introduction}

In quantum mechanics a system's behavior is not independent of how it is measured. As a result, the readout used in an experiment must be tailored to the phenomena of interest. Likewise, for a given type of readout not all quantum effects are observable.

Experiments on mechanical oscillators have to date used readouts that couple directly to the oscillator's displacement. The most common example is an optical interferometer in which the oscillator serves as one of the interferometer's mirrors. In such a system the phase $\phi$ of the light reflected from the interferometer is proportional to the mirror's displacement $x$. An oscillator that is subject to continuous monitoring of $x$ is predicted to show a number of striking quantum features, including the standard quantum limit of displacement detection.\cite{BraginskyKhalili} Additionally, the linear coupling between $x$ and $\phi$ can be used both to laser-cool the oscillator (perhaps eventually to its ground state)\cite{Marquardt,Wilson-Rae,Linewidth} and to squeeze the light leaving the cavity.\cite{Fabre,Tombesi} The connection between the readout of the mechanical oscillator and its manipulation highlights the fact that these are two aspects of the same optomechanical coupling.

In a recent paper\cite{nature} it was shown that a modest rearrangement of the usual optomechanical setup can realize a fundamentally different type of readout. When a nearly-transparent dielectric membrane is placed inside a cavity formed by two fixed, macroscopic mirrors, the phase of the light reflected from the cavity can be adjusted so that it is proportional either to $x$ or to $x^2$. The quadratic coupling occurs when the membrane is placed at a node (or anti-node) of the intracavity standing wave. In such a situation the membrane is at a minimum (maximum) of the optical intensity, and so detunes the cavity resonance by a small (large) amount. As the membrane moves in either direction it encounters an optical intensity that is larger (smaller) by an amount quadratic in its displacement (to lowest order), and hence detunes the cavity by an amount which is also quadratic (to lowest order) in the displacement. If on the other hand the membrane is originally placed at a point which is neither a node nor an antinode, the cavity detuning is (to lowest order) linear in the displacement.

Mechanical oscillators coupled to an $x^2$-readout have been discussed theoretically for some time. It has been shown that such a readout, coupled to a mechanical oscillator inside a sufficiently high-finesse optical cavity, can in principle provide a quantum nondemolition (QND) measurement of the energy (or equivalently the phonon number) of the mechanical oscillator.\cite{BraginskyScience1980} With a sufficiently sensitive $x^2$-readout it should be possible to observe, in real time, the individual quantum jumps of the mechanical oscillator. This is in contrast to an oscillator coupled to an $x$-readout, in which the repeated measurements of the oscillator's position extract information which prevents the oscillator from remaining in an energy eigenstate. This is because the quantity $x$ does not commute with the oscillator's energy, whereas the quantity $x^2$ does (at least in the rotating-wave approximation, whose validity is ensured by the cavity's high finesse).\cite{BraginskyScience1980}

Although the $x^2$-readout demonstrated in reference [7] represented a major advance towards realizing the goal of QND measurements of a mechanical oscillator's energy, the strength of the $x^2$-coupling was insufficient to realize such a measurement in practice. This is because for a low-reflectivity membrane the scale of the $x^2$-coupling is $\sim 1/\lambda^2$, where $\lambda$ is the wavelength of the light (the cavity detuning oscillates each time the membrane is displaced by $\lambda/2$). If the membrane's (field) reflectivity $r$ approaches unity, the finesse of the ``half-cavities'' on either side of the membrane begins to increase, and the curvature of the cavity detuning (and hence the strength of the $x^2$-coupling) increases, diverging for $r \rightarrow 1$.\cite{nature} However the technical challenges involved in combining a high reflectivity mirror and a high-quality mechanical oscillator into a single element have proven considerable, so it would be highly advantageous to find a strong $x^2$-coupling which does not require a high-reflectivity membrane. 

In this paper, we describe a new means for generating a strong $x^2$-coupling in this type of device. We show that the optical cavity's full spectrum of transverse modes contains many near-degeneracies, and that near these points the cavity's resonance frequencies display an avoided-crossing behavior as a function of the membrane displacement. This leads to a detuning proportional to $x^2$, but with a scale set by the symmetry-breaking aspects of the cavity/membrane geometry rather than the wavelength of light. We develop a perturbation theory that allows us to calculate the membrane-induced cavity detuning, and find that the $x^2$-coupling at these avoided crossings can be made orders of magnitude stronger than realized in earlier work. We compare these calculations to measurements and find quantitative agreement, indicating that the single-phonon QND measurements proposed in [7] may be feasible even with a low-$r$ membrane.

\section{Observed Effect of Membrane on Empty-Cavity Modes}
\label{measurements}

\begin{figure}
\begin{center}
\includegraphics[width=3in]{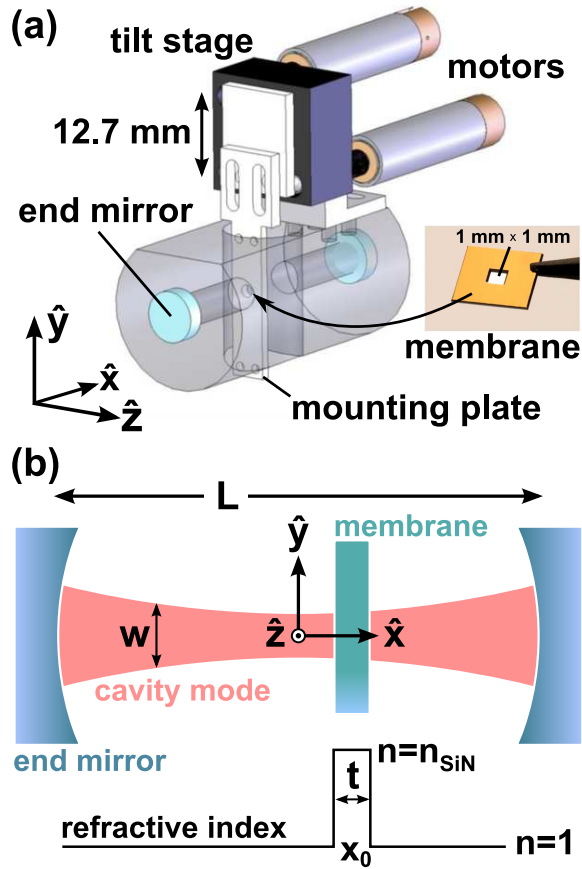}
\end{center}
\caption{(a) Schematic of our apparatus: A flexible SiN membrane mounted on a motorized tilt stage at the center of a Fabry-Perot cavity is coupled to the cavity's optical modes via radiation pressure. Piezoelectric actuators between the mounting plate and membrane enable displacements along the $x$-axis. (b) Simplified diagram of the cavity and membrane. The cavity length is $L=6.7$ cm and the end mirror radius of curvature is 5 cm.}
\label{geometry}
\end{figure}

Our experimental setup is shown in Fig. \ref{geometry} and has also been described elsewhere\cite{nature, NJP}. A flexible silicon nitride membrane (1 mm $\times$ 1 mm $\times$ 50 nm thick) is situated near the waist of a high-finesse Fabry-Perot cavity so that its normal vector is roughly parallel to the cavity's long ($x$) axis. The membrane acts as the micromechanical resonator and its deflection is coupled to the cavity's optical modes via radiation pressure. The two macroscopic end mirrors are held fixed by an Invar cavity spacer. A motorized tilt stage holding the membrane is mounted to the spacer, and two piezoelectric actuators are used to displace the membrane along $\hat{x}$.

\begin{figure}
\begin{center}
\includegraphics[width=4in]{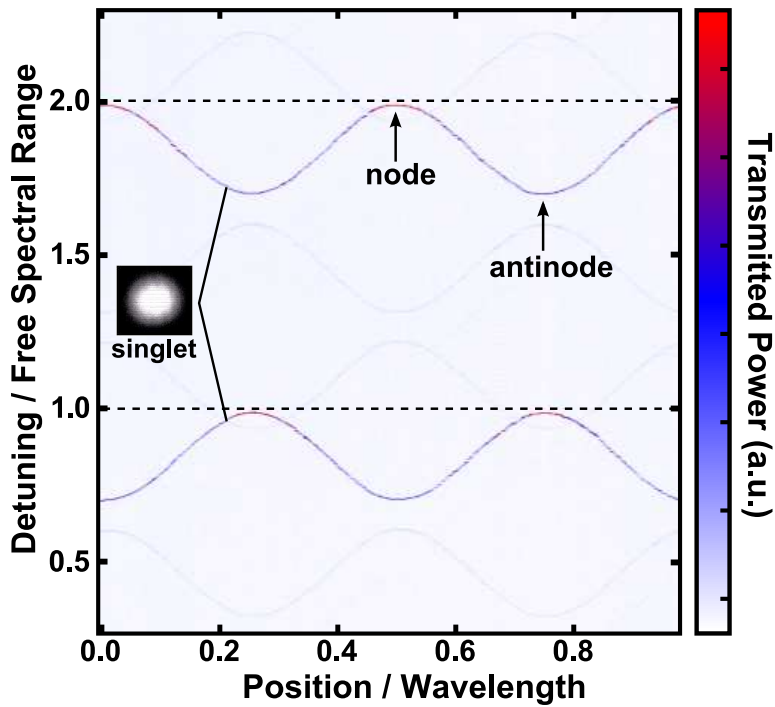}
\end{center}
\caption{Transmission through the cavity as a function of laser detuning and membrane position. The dominant signal corresponds to the TEM$_{0,0}$ (singlet) cavity mode. Dashed lines show the approximate position of the unperturbed singlet modes. We have labeled positions corresponding to a node and antinode of the upper singlet mode's electric field. At these points the detuning is proportional to $x^2$. (inset) An infrared camera image showing the transmitted beam profile.}
\label{bands}
\end{figure}

We can begin to characterize the optomechanical coupling in this system by measuring the transmission through the cavity as a function of membrane position and laser detuning, as shown in Fig. \ref{bands}. Here the laser is aligned so that the dominant transmission peak corresponds to the TEM$_{0,0}$ (singlet) mode, as confirmed by a camera monitoring transmission (inset).  As the membrane moves along the longitudinal ($x$) axis, it perturbs the cavity resonance frequencies to lower values, producing a detuning that varies roughly sinusoidally with position. 

When the membrane is located at an optical node, the perturbation is minimal, and the detuning is quadratic in position. As a result, light leaving the cavity contains only information about $x^2$. As discussed elsewhere\cite{nature}, this may enable QND phonon number readout using the TEM$_{0,0}$ mode alone. Since the membrane is a thin ($50$ nm) dielectric ($n_{\text{SiN}} \approx 2$), it is a very poor reflector ($|r|^2 = 0.13$ where $|r|^2$ is the power reflectivity). As a result the curvature of the detuning is small and the $x^2$-sensitivity is weak. Practical estimates predict that in order to observe a phonon Fock state before it decays, the membrane reflectivity would need to be substantially higher, $\sim 0.998$\cite{nature}. This may represent the most difficult of the technical challenges to observing real-time quantum jumps of the membrane's mechanical energy.

\begin{figure}
\begin{center}
\includegraphics[width=6in]{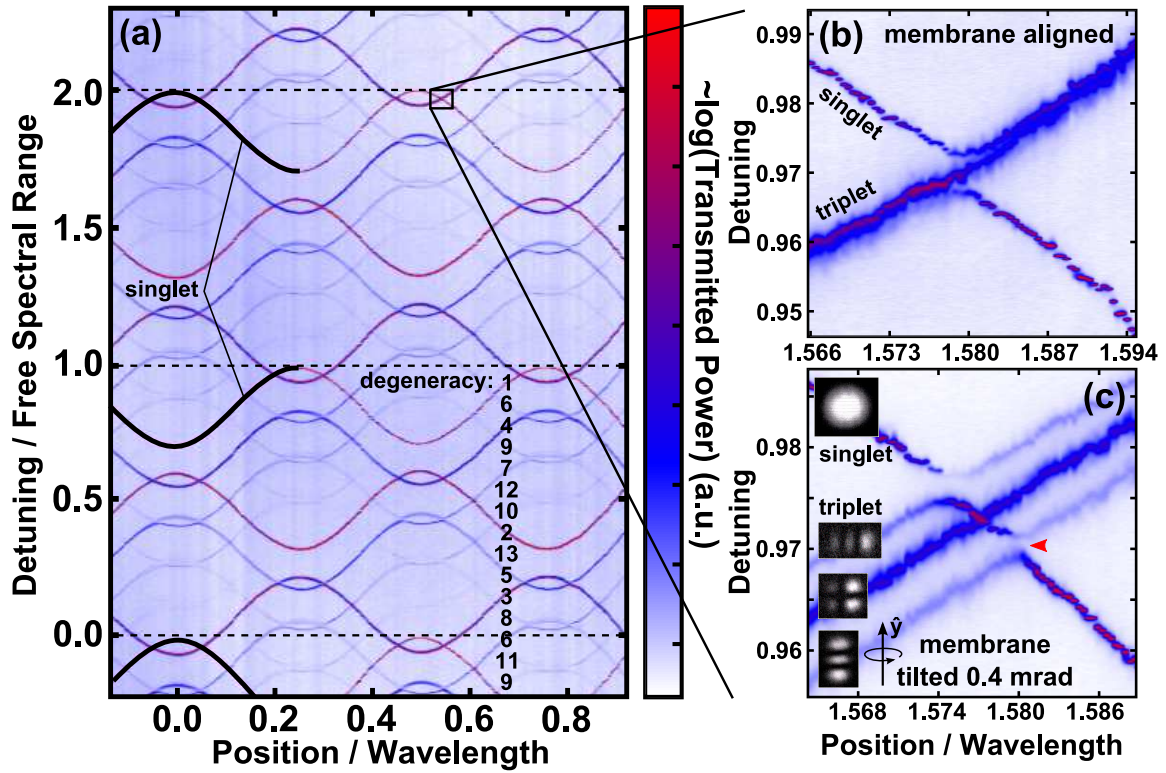}
\end{center}
\caption{(a) Transmission spectrum with the input laser misaligned, plotted on a log scale to enhance the faint features. The degeneracies of the different transverse modes are labeled and solid lines are drawn over the singlet mode for reference. (b) Close-up of the singlet-triplet crossing point for the membrane aligned with its normal vector parallel to the cavity axis ($\hat{x}$). (c) Singlet-triplet crossing with the membrane tilted about the $y$-axis by 0.4 mrad. The strength of the curvature at the marked gap corresponds to an effective membrane reflectivity of $0.994 \pm 0.001$ power.}
\label{crossings}
\end{figure}

A promising solution to this problem lies in the interactions between different transverse optical modes. We can couple to and identify many more of the cavity's transverse modes by intentionally misaligning the input laser, as shown in Fig. \ref{crossings}(a). We have identified all of the visible bands, such as the \{TEM$_{1,0}$, TEM$_{0,1}$\} doublet, \{TEM$_{2,0}$, TEM$_{1,1}$, TEM$_{0,2}$\} triplet, and so on up to the 13-fold degenerate (tridectet) modes.

The different transverse modes cross each other as a function of position at several places in Fig. \ref{crossings}. Figure \ref{crossings} also shows a close-up of the crossing between the singlet and the triplet with the membrane's normal vector (b) aligned with $\hat{x}$ (c) tilted around the $y$-axis by 0.4 mrad. Tilting the membrane as in (c) lifts the degeneracy of the triplet in a predictable way: modes extended the furthest in the $\hat{z}$ direction shift the most. As is evident from Fig. \ref{crossings}(c), the crossing points between the singlet and the two \emph{even} triplets (TEM$_{2,0}$ and TEM$_{0,2}$) are avoided, meaning that in addition to perturbing the individual modes, the membrane also couples them.

Most importantly, the quadratic detuning at the avoided crossing turning points is very strong. In Fig. \ref{crossings}(c), the curvature is 50 times stronger than at the single-mode turning points. This is the same curvature a membrane reflectivity of $0.994 \pm 0.001$ would generate using a single mode \cite{nature}. This is already very close to the QND target and we have not yet attempted to optimize the system.\footnote{As discussed later, we have observed smaller gaps, but for that data the fit curvature is not very convincing due to vibrations limiting our displacement sensitivity. We can still use these smaller gaps to infer a lower bound on the curvature.}

\section{Model}

While the membrane perturbs the empty cavity modes by up to a quarter of a free spectral range, its fractional effect relative to the laser frequency is minute ($\sim 10^{-6}$). We can therefore view the membrane perturbatively and develop a first-order theory to model the system, as discussed in the next section. We then outline a method by which to solve this problem analytically when the membrane is positioned near the cavity waist, and finally compare our results with measurements. 

\subsection{First-Order Degenerate Perturbation Theory}

We start with the time-independent free-space electromagnetic wave equation
\begin{equation}
\nabla^2 \phi + \frac{\omega^2}{c^2} \phi = 0
\label{freespace_wave}
\end{equation}
where $\omega$ is the angular frequency and $c$ is the speed of light. As is drawn in Fig. \ref{geometry}(b), we define the origin to reside at the center of our cavity with the $x$-axis pointing toward one of the (spherical) end-mirrors. Under these boundary conditions, a convenient set of (Hermite-Gaussian) orthonormal solutions is given by \cite{siegman}
\begin{eqnarray}
\phi_j &=& \frac{H_m(\sqrt{2}y / w) H_n(\sqrt{2} z / w)}{w \sqrt{\pi L 2^{m+n-1} m! n!}} e^{-(y^2+z^2)/w^2} \nonumber\\
			 &\times& e^{i(m+n+1)\Psi} e^{-i k (y^2+z^2) / 2 R} e^{-i k x-i l \pi/2},
\label{empty_modes}
\end{eqnarray}
Here $H_m$ is the $m^{\text{th}}$ Hermite polynomial, $k = \omega/c$ is the wavenumber, $w(x)=\sqrt{2(x^2+x_R^2)/k x_R}$ is the width of the cavity mode at $x$ (where $x_R = 2.351$ cm is the Raleigh range and $w_0 = 89.2~\mu$m is the waist for our geometry), $L = 6.7$ cm is the cavity length, $m$ and $n$ are the transverse mode indices, $l$ is the longitudinal mode index, $\Psi(x)=\tan^{-1}(w^2 k/2 R)$ is the Guoy phase shift, and $R(x)=(x^2+x_R^2)/x$ is the wave fronts' radius of curvature. For a standing wave in our cavity, $\text{Re}(\phi_j)$ is proportional to the electric field amplitude, and the $l \pi/2$ term ensures that each longitudinal mode's electric field is zero at the end mirrors. The prefactors ensure that the inner product $\int dx\int dy \int dz \text{Re}(\phi_i) \text{Re}(\phi_j) = \delta_{ij}$. 

As shown in Fig. \ref{geometry}(b), we represent the membrane as a block of refractive index $n_{\text{SiN}} \approx 2$ and thickness $t=50$ nm, centered at position $x_0$. This modifies the speed of light in this short region, so that the wave equation in the cavity becomes
\begin{equation}
\nabla^2 \psi + \frac{\omega^2}{c^2}(1 + V(x-x_c)) \psi = 0
\label{perturbed_wave}
\end{equation}
where $V(x-x_c) = (n_{\text{SiN}}^2-1)\left(\Theta[x-(x_c-t/2)] - \Theta[(x_c+t/2)-x]\right)$ and $\Theta$ the Heaviside step function. For an ``aligned'' membrane (i.e. flat in the $y$-$z$ plane) $x_c = x_0$ is constant. To incorporate tilt into the model, let $x_c = x_0 + \alpha_y y + \alpha_z z$ where $\alpha_y$ and $\alpha_z$ are the small rotations about the $z$ and $y$ axes, respectively. 

\begin{figure}
\begin{center}
\includegraphics[width=3in]{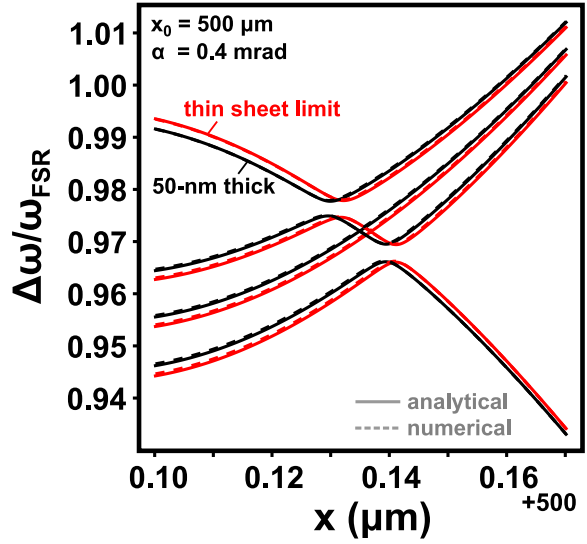}
\end{center}
\caption{Comparison of numerical (dashed) and analytical (solid) results near the singlet-triplet crossing points. The red curves correspond to the thin-membrane (delta function) limit, and the black lines include membrane thickness. For this plot, $x_0$ = 500 $\mu$m, $\alpha_z$ = 0.4 mrad, and $\alpha_y=0$.}
\label{numerical}
\end{figure}

The perturbed modes $\psi$ can be expanded in terms of the empty-cavity modes:
\begin{equation}
\psi = c_1 \phi_1 + c_2 \phi_2 + c_3 \phi_3 + ...
\end{equation}
where the $c$'s are constants. We wish to study the region of near-degeneracy shown in Fig. \ref{crossings}(b-c), between the $l$-th longitudinal singlet mode ($m=n=0$) and the three $(l-1)$-th triplet modes ($m+n=2$), so we make the assumption that $\psi$ is composed mostly of these four empty-cavity modes
\begin{equation}
\psi = c_s \phi_s + c_y \phi_y + c_a \phi_a + c_z \phi_z + \sum{\epsilon_j \phi_j}
\end{equation}
where the indices $s$, $y$, $a$, and $z$ refer to the singlet, the triplet widest in the $\hat{y}$ direction ($m=2,n=0$), the antisymmetric triplet ($m=1,n=1$), and the triplet widest along $\hat{z}$ ($m=0,n=2$), respectively. The last term is a summation over all remaining modes, and its contribution is assumed to remain small ($\epsilon_j \ll 1$). We also assume $\psi$ will have a new eigenvalue $\omega^2/c^2 \equiv \kappa$ that is not very different (i.e. within a fraction of a free spectral range) from any of the unperturbed eigenvalues of the four contributing $\phi$'s. Substituting this into Eq. \ref{perturbed_wave},
\begin{equation}
(\nabla^2 + (1+V) \kappa) (c_s \phi_s + c_y \phi_y + c_a \phi_a + c_z \phi_z + \sum{\epsilon_j \phi_j}) = 0.
\end{equation}
If we now take an inner product of this equation with each of the four empty-cavity modes and divide through by $\kappa$, we obtain four new equations
\begin{equation}
(1 - \kappa_i/\kappa) c_i + V_{is} c_s + V_{iy} c_y + V_{ia} c_a + V_{iz} c_z + \sum V_{ij}\epsilon_j = 0
\label{inner_product}
\end{equation}
where the index $i$ is $s, y, a,$ or $z$ and $V_{ij}$ is the inner product of the $i$-th and $j$-th mode with $V(x-x_c)$. The inner products $V_{ij}$ involve an integral over thickness $t$ between two modes that are normalized over length $L$ and are small (of order $(n_{\text{SiN}}^2-1)t/L \sim 2 \times 10^{-5}$ or less), so the last term in Eq. \ref{inner_product} can be ignored. We can further simplify by writing $\kappa \equiv \kappa_s (1+\delta)$ and $\kappa_{y,a,z} \equiv \kappa_s (1+g)$ where $\delta$ is the fractional change due to the membrane and $g$ is the (constant) fractional separation of the unperturbed singlet and triplet bands due to the Guoy phase. Both $\delta$ and $g$ are of order $10^{-5}$. To first order, the remaining equation can be written as a matrix
\begin{equation}
\left( \begin{array}{cccc}
\delta  +V_{ss} & V_{sy} & V_{sa} & V_{sz} \\
V_{sy} & \delta-g+V_{yy} & V_{ya} & V_{yz} \\
V_{sa} & V_{ya} & \delta-g+V_{aa} & V_{az} \\
V_{sz} & V_{yz} & V_{az} & \delta-g+V_{zz}\end{array} \right)
\left( \begin{array}{c}
c_s\\
c_y\\
c_a\\
c_z\end{array}\right) = 0.
\label{matrix}
\end{equation}
Solving this eigenvalue problem for $\delta$ in terms of $g$ and the $V$'s is straightforward and, though time-consuming, it is also easy to numerically compute $V_{ij}$. The problem is in principle solved, and the result of such a calculation is shown in Fig. \ref{numerical}. Computation time can also be reduced by assuming the membrane is an infinitesimally thin sheet (also plotted), but even a small finite thickness of $t = 50$ nm produces a noticeable effect.

\subsection{Analytical Solution Near the Cavity Waist}

We can also solve the inner products $V_{ij}$ analytically by using an approximate form for the unperturbed modes near the cavity waist. We do this by expanding $1/R(x)$, $w(x),$ and $\Psi(x)$ in Eq. \ref{empty_modes} in terms of the small parameter $\Delta = x/x_R$:
\begin{eqnarray}
\frac{1}{R(x)} &\approx& \frac{\Delta}{x_R} + O^3              \\
w(x)           &\approx& w_0 (1+\frac{1}{2}\Delta^2) + O^3     \\
\Psi(x)        &\approx& \Delta + O^3
\label{expansion}
\end{eqnarray}
Substituting this into Eq. \ref{empty_modes} yields
\begin{eqnarray}
\phi_j &=& \frac{H_m(\sqrt{2}y/w) H_n(\sqrt{2} z/w)}{w \sqrt{\pi L 2^{m+n-1} m! n!}} e^{-(y^2+z^2)/w^2}  \nonumber\\
			 	&\times& e^{i\left[(n+m+1-\frac{y^2+z^2}{w^2}-k x_R)\Delta + l \pi / 2\right]}.
\label{approx_modes}
\end{eqnarray}
We can now use this to estimate the inner products $V_{ij}$. We will not specify which modes are under consideration, so the results of this section may be applied to any set of nearly-degenerate cavity modes. 

Before we attempt to solve these integrals, first note that
\begin{equation}
\iiint \text{Re}(\phi_i) V \text{Re}(\phi_j) = \text{Re}\left[\frac{1}{2}\iiint \phi_i \phi_j V + \frac{1}{2}\iiint \phi_i \phi_j^* V\right]
\label{two_integrals}
\end{equation}
which simplifies the calculation. The first integral is by far the most challenging and so we outline its solution here.

First, note that the singlet and triplet modes have slightly different unperturbed $k$'s and $w$'s. If $i=s$ and $j \in \left\{y,a,z\right\}$, then $k_j$ = $k_i(1+g)$ and $w_{0,j} \approx w_{0,i}(1-g/2)$. By defining $A \equiv 1+g/2 \sim 1$, we can easily keep track of this difference to first order (and $A=1$ if the modes belong to the same degenerate manifold). If we plug Eq. \ref{approx_modes} into the first integral of Eq. \ref{two_integrals}, make the substitutions $y \rightarrow w y / \sqrt{2 A}, z \rightarrow w z \sqrt{2 A}$ and $x \rightarrow x_R \Delta$, we have
\begin{eqnarray}
&&\frac{1}{2}\iiint \phi_i \phi_j V = \frac{P_0}{\Delta_t} e^{-i (l_i+l_j) \pi/2} \nonumber\\
&&~~~~\times \int dy \int dz \int_{\Delta_c-\Delta_t/2}^{\Delta_c+\Delta_t/2} d\Delta ~p(y,z) e^{-y^2-z^2} e^{-i(K + y^2 + z^2)\Delta}
\label{first_integral}
\end{eqnarray} 
with
\begin{eqnarray}
P_0      &=& \frac{(t/\cos \alpha)(n_{\text{SiN}}^2-1)}{\pi L \sqrt{2^{m_i+n_i+m_j+n_j} n_i! m_i! n_j! m_j!}} \\
\Delta_t &=& t / x_R \cos \alpha																				\\
\Delta_c &=& x_c/x_R = \Delta_0+\beta_y y+\beta_z z	\\
p(y,z)   &=& H_{m_i} (y/\sqrt{A}) H_{m_j}(\sqrt{A} y) H_{n_i} (z/\sqrt{A}) H_{n_j}(\sqrt{A} z)								\\
K        &=& 2 A k_i x_R - (m_i+n_i+m_j+n_j+2).
\end{eqnarray} 
Here $\Delta_t$ is the dimensionless membrane thickness corrected for tilt $\alpha = \sqrt{\alpha_y^2+\alpha_z^2}$, and we have allowed the position of the membrane center $x_c$ to depend on $y$ and $z$ through small (rescaled) tilts in both directions, $\beta_{y,z} = \alpha_{y,z} w_i / x_R \sqrt{2A}$.

The membrane is 20 times thinner than the free-space wavelength, but it noticeably affects the cavity modes, as is evident in Fig. \ref{numerical}. We can approximate the integral over $\Delta$ by noting that for a smooth function $f(x)$,
\begin{equation}
\int_{x_0-\delta x/2}^{x_0+\delta x/2} f(x) dx = \delta x f(x_0) +  \delta x^3 \frac{1}{24} f''(x_0) + O(\delta x^5)
\label{integral_expansion}
\end{equation}
Applying this to our integral,
\begin{eqnarray}
\frac{1}{2}\iiint \phi_i \phi_j V \approx P_0 T e^{-i (l_i+l_j) \pi/2} & \iint &dy~dz~ p(y,z) e^{-y^2} e^{-z^2} \nonumber\\
&\times &e^{-i\left[(\Delta_0+\beta_y y + \beta_z z)(y^2+z^2)\right]} \nonumber\\ 
&\times &e^{-i\left[(\Delta_0+\beta_y y + \beta_z z)K\right]}
\label{integral_no_z}
\end{eqnarray}
with $T = 1-\Delta_t^2 K^2/24$. It should be noted that $K \sim 250,000$ is very large, and so when estimating the thickness correction $T$ from Eq. \ref{integral_expansion}, we ignored several terms smaller than $\Delta_t^2 K^2 / 24 \sim 10^{-3}$ by a factor of $K$ or more. 

The exponent in the second line of Eq. \ref{integral_no_z} contains only small quantities, so we can simplify this term by making the expansion $e^{i\epsilon} \approx 1+i\epsilon$. Then we complete the square for $y$ and $z$ in the remaining exponential and make the variable change $y \rightarrow y-i\beta_y K/2$ and $z \rightarrow z-i\beta_z K/2$. If we then define the (analytically soluble) integral
\begin{equation}
\xi_{n_i n_j}^{q \beta K} = \int dx~(x-i\beta K/2)^q e^{-x^2} H_{n_i}(\frac{x-i\beta K/2}{\sqrt{A}})H_{n_i}((x-i\beta K/2) \sqrt{A})
\end{equation}
and the shorthand $\Gamma_{qp} \equiv \xi_{m_i m_j}^{q \beta_x K} \xi_{n_i n_j}^{p \beta_y K}$, Eq. \ref{integral_no_z} becomes 
\begin{eqnarray}
&&\frac{1}{2}\iiint \phi_i \phi_j V \approx P_0 T e^{-iK\Delta_0-i(l_i+l_j)\pi/2} e^{-\frac{K^2(\beta_x^2 + \beta_y^2)}{4}} \nonumber\\
&&~~~~\times \left[ \Gamma_{00} - i\left((\Gamma_{20}+\Gamma_{02})\Delta_0+(\Gamma_{30}+\Gamma_{12})\beta_y+(\Gamma_{03}+\Gamma_{21})\beta_z\right)\right].
\label{first}
\end{eqnarray}
Applying a similar method to the second half of Eq. \ref{two_integrals}, and with definitions $K' \equiv m_j-m_i+n_j-n_i-g k_i x_R$ (much smaller than $K$) and $\Gamma_{qp}' \equiv \xi_{m_i m_j}^{q \beta_x K'} \xi_{n_i n_j}^{p \beta_y K'}$, it can be shown that
\begin{eqnarray}
 \frac{1}{2}\iiint \phi_i \phi_j^* V & \approx & P_0 e^{-iK\Delta_0-i(l_i-l_j)\pi/2}        \nonumber\\
 & \times & \left[ \Gamma_{00}' - i K' \left(\Gamma_{0}'\Delta_0+\Gamma_{10}'\beta_y+\Gamma_{01}'\beta_z\right)\right].
\label{second}
\end{eqnarray}
Equations \ref{first}, \ref{second} and \ref{two_integrals} represent a very accurate analytical approximation of the inner products $V_{ij}$ for small displacements (relative to $x_R$) from the cavity waist. These results are also plotted (with and without the thickness correction) in Fig. \ref{numerical} of the previous section as solid lines. In practice, the agreement for our setup is excellent as long as $|x_c| < 1$ mm. As expected, the approximation is nearly perfect at the waist and breaks down as $x_0$ approaches $x_R$.

\subsection{Discussion and Comparison with Data}

We can gain insight into our system from the analytical results. By ignoring the off-diagonal terms in Eq. \ref{matrix} (i.e. ignoring avoided crossings), the eigenvalues simplify substantially, and the detuning is given by
\begin{eqnarray}
\frac{\Delta \omega_s}{\omega_0}       &\approx& -\frac{t(n_{\text{SiN}}^2-1)}{2L} (1 - T \cos((2k_s x_R - 1) \Delta_0))  \\
\frac{\Delta \omega_{y,a,z}}{\omega_0} &\approx& -\frac{t(n_{\text{SiN}}^2-1)}{2L} (1 + T \cos((2k_{y} x_R - 3) \Delta_0)) + g
\label{phase}
\end{eqnarray}
for the singlet and triplet modes respectively. These equations represent two sinusoidal bands oscillating (with opposite sign) below their unperturbed detunings, with peak-to-peak amplitudes $T(n_{\text{SiN}}^2-1)t/L \approx 27\%$ of the free spectral range, and separated from each other by the appropriate Guoy spacing. Applying this method to the other transverse modes, we can generate the entire band structure shown in Fig. \ref{crossings}(a), and for $n_{\text{SiN}}=2.0$ the agreement is essentially perfect.

The spatial period of the singlet band is slightly smaller than the triplet, and so it should oscillate a little faster as a function of $\Delta_0$. Though subtle, we do observe a phase difference between the bands (this is somewhat more visible in Fig. \ref{crossings} when comparing the singlet and the nonet band), and we can use this phase difference to estimate the membrane's displacement $x_0$. In the time it takes to raster the band structure in Fig \ref{crossings}, drifts in the piezos and laser frequency cause noticeable distortions of the bands on the scale of this phase difference, but nonetheless by fitting the neighboring singlet and triplet sinusoids from Fig. \ref{crossings}(a) we estimate a positive (as defined by the $x$-axis) displacement of $300 \pm 100~\mu$m from the waist. 

It is also worth noting that the effect of finite membrane thickness is to wash out the oscillations (i.e. $T$ decreases from unity as $t$ increases). This makes sense qualitatively because when the membrane is positioned at a node, the electric field is not zero everywhere inside it and so there will still be a small negative perturbation at the top of the band. Similarly, when the membrane is at an antinode, the field is not maximal everywhere inside it and so the perturbation is not as strong. Optical losses inside the membrane mean that even when positioned at a node the finite thickness will put an upper bound on the finesse this system can achieve. We have, however, already observed a finesse of 150,000 with the membrane inside the cavity \cite{NJP}. 

If we now look at the mode-coupling terms $V_{i\neq j}$, we can gain some insight into the avoided crossings of Fig. \ref{crossings}(b-c). First, all of the off-diagonal terms involving the antisymmetric mode (TEM$_{1,1}$) are identically zero (even with $\alpha \neq 0$) in this approximation. This is a reflection of the fact that the TEM$_{1,1}$ mode is an odd function in both the $y$ and $z$ directions, while the other three modes are even. The integrals across the membrane therefore all involve a function that is approximately odd and vanish. Hence there should be no avoided crossing between TEM$_{0,0}$ and TEM$_{1,1}$ to first order, which agrees with all of our observations (see Fig. \ref{crossings}(c), for example).

The other off-diagonal terms are not zero (thankfully), and the result is again relatively simple if we keep the membrane aligned (i.e. $\alpha=0$). 
\begin{equation}
V_{sy} = V_{sz} \approx -\Delta_0 \frac{t(n_{\text{SiN}}^2-1)}{2L} T \cos\left[((k_s+k_z)x_R-4)\Delta_0\right].
\label{interaction}
\end{equation}
When the membrane is aligned, the interaction is proportional to $\Delta_0$ times a term that oscillates with a period close to that of the bands (though since the bands always cross each other at roughly the same phase, this term will modulate the coupling slowly as a function of $\Delta_0$). Following this backward through the calculation in the previous section, we see that it arises from our expansion of the finite radius of curvature $R$. So (perhaps not surprisingly) the interaction between these modes arises from the mismatch between the curved wavefronts and the flat membrane.\footnote{Naturally, if the membrane distorted to follow a constant phase front as it moved from the waist, the different transverse modes would all remain orthogonal. Perhaps another way to think of the mismatch-induced coupling is to imagine a curved wave partially reflecting from a flat surface. It will certainly not scatter entirely back into the same mode.}

\begin{figure}
\begin{center}
\includegraphics[width=5.5in]{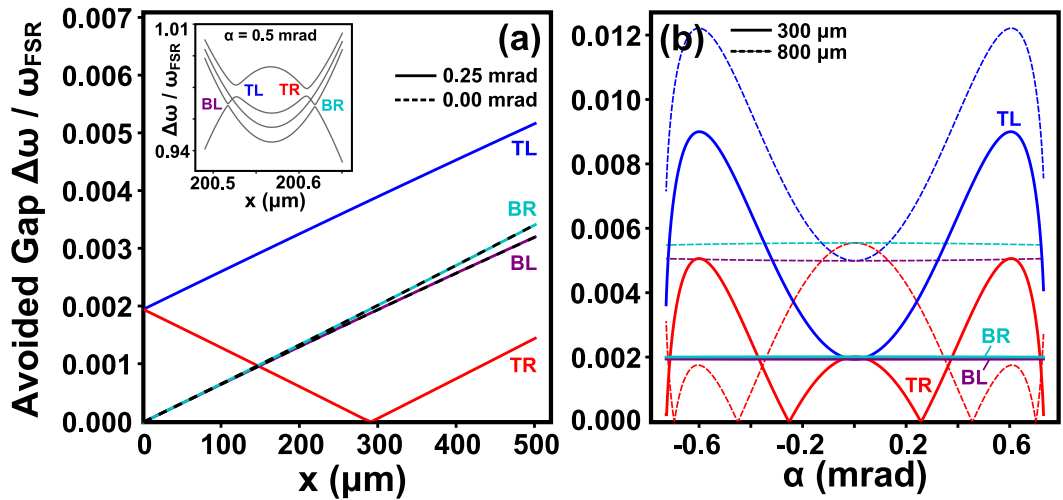}
\end{center}
\caption{(a) Dependence of the four avoided crossing gaps on membrane position, fixing tilt at 0 (dashed lines) and 0.25 mrad. (inset) Plot of the mode detuning versus position near the singlet-triplet crossings. The four gaps in (a, inset) are labeled for reference. (b) Dependence of the four gaps on membrane tilt, fixing the position at 300 and 800 $\mu$m. The TR gap should be adjustable over a wide range.}
\label{tunable}
\end{figure}

This is an encouraging result because it implies a strong degree of tunability in the avoided gap and hence the $x^2$-sensitivity. Figure \ref{tunable}(a) shows our calculation of the gaps at each of the four avoided crossings as a function of displacement out to 500 $\mu$m from the waist. In this plot we show results for both the aligned case (dashed lines) and for a tilt of 0.25 mrad (solid lines). For the aligned case, the gaps collapse onto two similar curves (as they must by symmetry), and when the membrane is tilted, the top two gaps (TR and TL of the inset) move to larger initial values.\footnote{Similar intuition applies here. The flat membrane no longer encloses a constant phase front.} The top right (TR) gap is a more interesting function of position, as it is tunable through zero at finite offset. If instead we fix the position of the membrane, we should also be able to tune the TR gap over a wide range (including zero) with tilt, as shown in Fig. \ref{tunable}(b). Note the large quantitative and qualitative differences between the TR and TL gaps can help calibrate the magnitude as well as the direction of the membrane displacement if it is not already known.

\begin{figure}
\begin{center}
\includegraphics[width=3.5in]{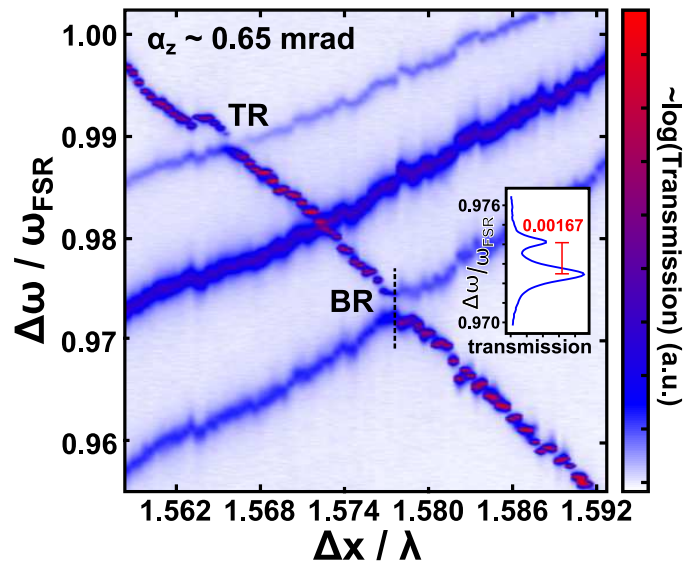}
\end{center}
\caption{Transmission data at large membrane tilt, $\alpha_z \approx$ 0.65 mrad. The dashed line corresponds to the data shown in the inset. (inset) Single-shot measurement of transmission versus laser detuning.}
\label{smallgap}
\end{figure}

As mentioned in section \ref{measurements}, vibrations in this apparatus preclude reliable determinations of gap size and $x^2$-sensitivity for very small gaps. Nonetheless we have observed some smaller gaps as we tune the membrane tilt, two of which are shown in Fig. \ref{smallgap}. In this data set, displacement noise due to ambient vibrations, coulomb forces on the membrane and/or piezo noise is quite evident. Further, during the left-to-right rastering of these data sets (acquired over $\sim 10-20$ minutes each) the laser temperature varied enough to cause a systematic detuning and sheer the data vertically. It is a large effect in this data set, and we are studying ways to compensate for it. 

In a given frequency sweep (vertical trace), however, the time it takes to traverse one of these gaps is roughly a millisecond; such a single-shot measurement of the mode spacing should therefore be much less susceptible to vibrations and drift. If we therefore record the smallest spacing in Fig. \ref{smallgap}, we can put a lower bound on the detuning curvature using the form detuning takes near an avoided crossing, $\sqrt{(a x)^2+(\Delta f/2)^2}$ where $a$ is the asymptotic slope and $\Delta f$ is the gap (both of which we estimate from Fig. \ref{smallgap}). Doing so yields a lower bound on the effective membrane reflectivity of $|r|^2 > 0.992 \pm 0.004$ for the TR crossing and $|r|^2 > 0.9989 \pm 0.0005\%$ for the BR crossing (the sharper curvature of BR reflects the larger asymptotic slope $a$). This estimate is still subject to vibrations above a few kilohertz, which we have not characterized. On the other hand, when we fit the curvature explicitly as in section \ref{measurements}, even lower-frequency vibrations (i.e. anything above about 0.1 Hz) can wash out sharp curvature, so that technique represents a very conservative estimate.

\begin{figure}
\begin{center}
\includegraphics[width=4in]{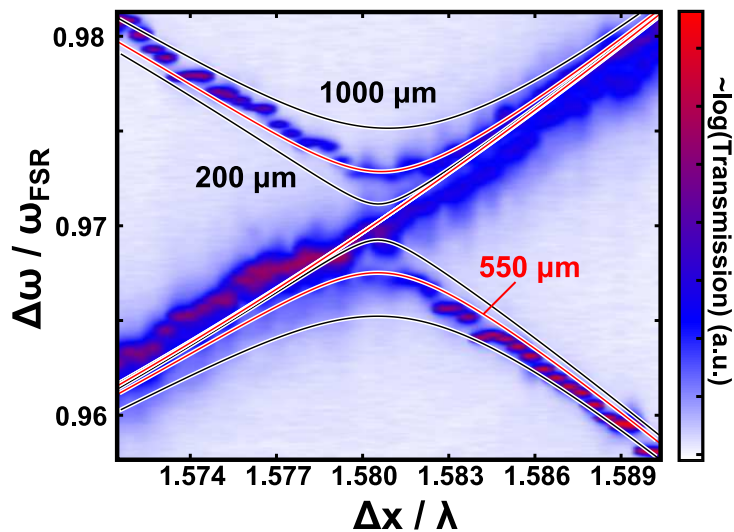}
\end{center}
\caption{Analytical model plotted on top of transmission data for the aligned membrane. Here we show the analytical results for the membrane situated at 200, 550, and 1000 $\mu$m from the cavity waist. }
\label{alignedoverlay}
\end{figure}

Figure \ref{tunable} implies we can use the TR and BR gaps to estimate the position of our membrane relative to the waist. Figure \ref{alignedoverlay} shows the aligned singlet-triplet crossing data from Fig. \ref{crossings}(b) along with curves generated by this model (for $\alpha=0$) at several different membrane positions. We estimate the membrane's displacement from the waist to be about $550~\mu$m here, which is a reasonable value for our apparatus and is in rough agreement with our previous estimate based on the horizontal offsets in the various bands shown in Fig. \ref{crossings}.

\begin{figure}
\begin{center}
\includegraphics[width=4.5in]{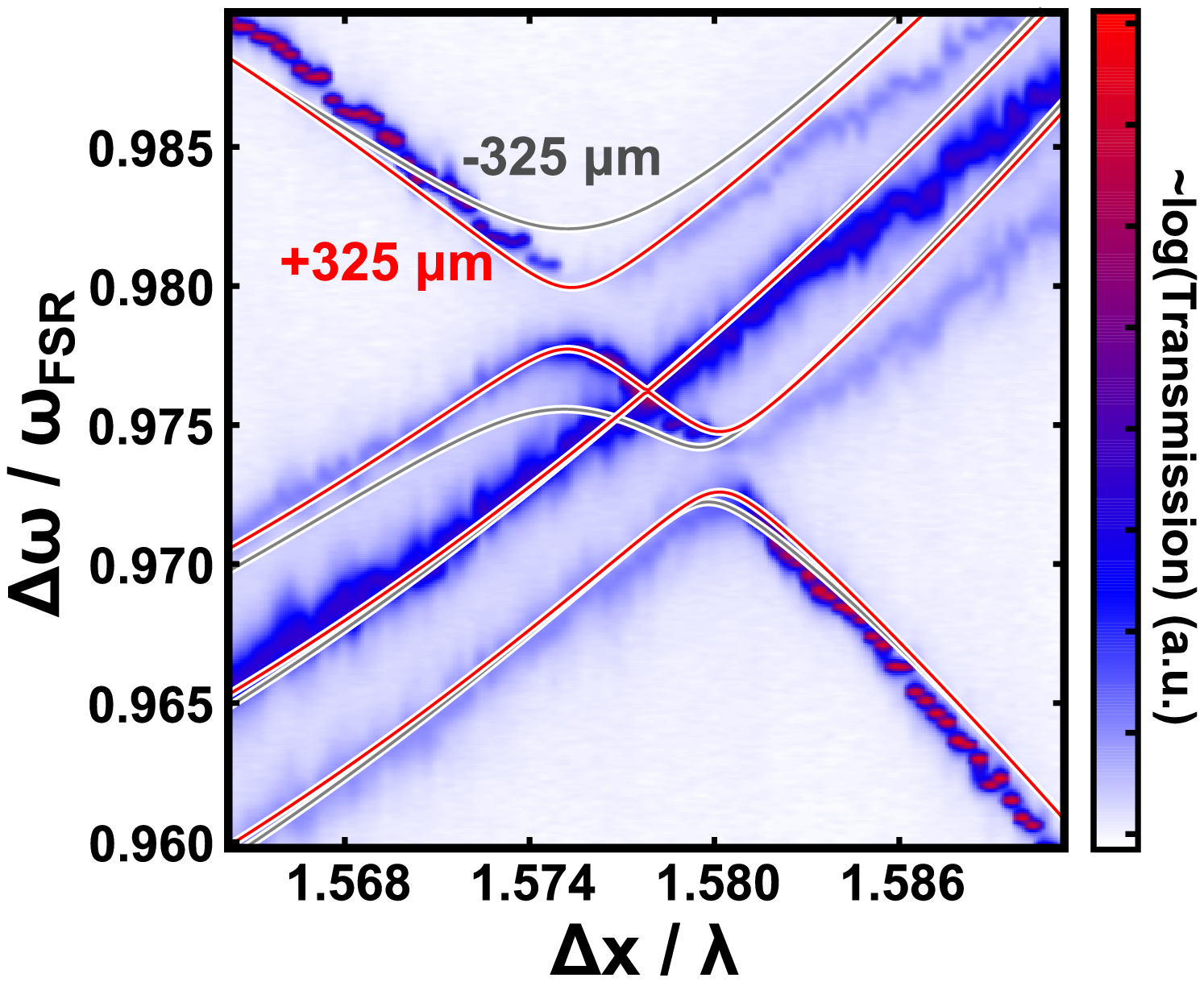}
\end{center}
\caption{Model plotted on top of the transmission data for the tilted membrane. Here we show the analytical results for the membrane situated at +325 and -325 $\mu$m from the cavity waist. }
\label{tiltedoverlay}
\end{figure}

We can further check the model for consistency by studying the interplay between tilt and the lifting of triplet degeneracy far from a crossing. It is relatively straightforward to show that this scales as $\alpha^2$ for small $\alpha$. The triplet splitting should also be quite insensitive to membrane position so we can use it to estimate the membrane's true tilt or even align the membrane.\footnote{This is in fact how we determined $\alpha\approx0$.} Figure \ref{tiltedoverlay} shows the data from Fig. \ref{crossings}(c) along with the analytical result for $x_0 = 325~\mu$m and $\alpha_z$ = 0.395 mrad. We obtain these parameters by first adjusting $\alpha_z$ until the triplet splitting is correct and then varying $x_0$ to match the avoided crossings. We have also plotted the result for displacement in the opposite direction, which essentially amounts to comparing our data with BL and TL in Fig. \ref{tunable}. The fit does not agree with the data here or at any other negative value of $x_0$. The smaller TR gap therefore confirms the sign of our membrane displacement (and it is the primary reason we chose to study the right side).

\begin{figure}
\begin{center}
\includegraphics[width=3in]{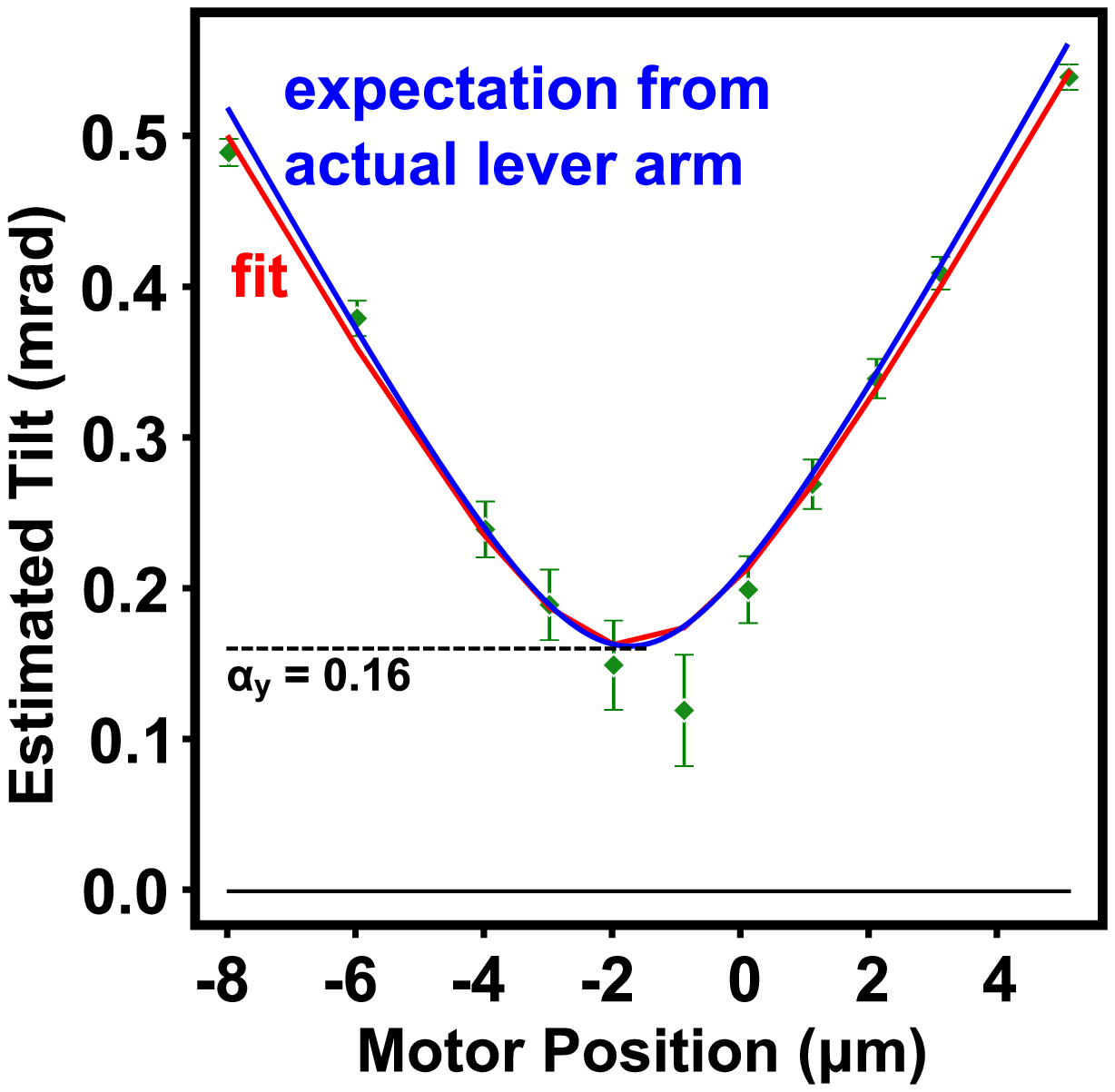}
\end{center}
\caption{Plot of tilt estimated from triplet splitting versus our tilt stage motor position. The red curve is a fit allowing the motor's linearity constant and misalignment $\alpha_y$ to float and the blue curve is the expectation for the same $\alpha_y$ and determining the linearity constant from the tilt stage geometry. }
\label{motor}
\end{figure}

We have performed similar analysis at several different values of the tilt stage's motor position, and these are summarized in Fig. \ref{motor}. Assuming there is a small constant tilt $\alpha_y$, we can fit this data with the form $\alpha = \sqrt{(a q_z)^2 + \alpha_y^2}$ where $a$ is a mechanical conversion factor between motor position $q_z$ and tilt. From the fit $\alpha_y = 0.16 \pm 0.01$ mrad and $a = 0.0756 \pm 0.0001$ mrad/$\mu$m. From the length of the tilt stage lever arm (12.7 mm) alone we estimate $a = 0.0787$ mrad/$\mu$m, implying a calibration error of $\sim 4\%$. We have plotted the expected result for the same $\alpha_y$ using a 12.7-mm lever arm for reference.

It is also important to note here that in the model we reproduce the ordering of the triplet modes: as we rotate the membrane about the $y$ axis, the modes most extended in the $z$ direction move the furthest. 

The model is in reasonable agreement with the data thus far, and it implies that in future experiments we should be able to tune the $x^2$-sensitivity to essentially any desired value. This could be a very important tool in our attempt to perform QND measurements of a single phonon.

\section{Summary/Outlook}

In this paper we have demonstrated that a SiN membrane can couple two nearly-degenerate transverse optical cavity modes, generating an avoided crossing and a cavity detuning that is strongly quadratic in membrane displacement $x$. Without optimizing the system, we have shown that this $x^2$-dependence (which is tunable over a wide range via membrane tilt) can be as strong as that generated using a single cavity mode and a membrane of reflectivity $|r|^2 \geq 0.9989 \pm 0.0005$. This means it might still be possible to perform QND measurements of phonon number in a membrane of modest reflectivity (i.e. $|r|^2 \sim 0.13$). We also derived a perturbative model of the system that quantitatively agrees with observations and further predicts the $x^2$-strength should be tunable to arbitrary strength through mm-scale membrane displacements.

These results should be taken with the caveat that the sharp avoided crossings described above occur when the membrane is not at a node of the intracavity field. As discussed previously\cite{nature,NJP}, this means that the optical loss in the membrane will limit the maximum cavity finesse. Whether or not the effect of this reduced finesse can be offset by the very strong quadratic coupling or reduced optical loss (e.g., via improved membrane materials or further engineering of the cavity modes) remains to be seen.

\end{document}